\documentstyle[11pt,epsfig,here]{article}
\begin{document}
\begin{center}
{\Large
{\it
Submitted to the Proceedings of 1st Intern. Conference on\\[1pt]
Non Accelerator New Physics NANP-97\\[1pt] 
(Russia, Dubna, July 7 -- 11, 1997)\\[1pt]
}}
\end{center}

\vspace{11mm}

\begin{center}
{\LARGE \bf The Baikal Deep Underwater Neutrino Experiment:}

{\LARGE \bf Status Report} 
\end{center}

\begin{center}
THE BAIKAL COLLABORATION:

V.A.Balkanov$^2$, I.A.Belolaptikov$^7$, L.B.Bezrukov$^1$, N.M.Budnev$^2$, 
A.G.Chensky$^2$, I.A.Danilchenko$^1$, Zh.-A.M.Djilkibaev$^1$, 
G.V.Domogatsky$^1$, A.A.Doroshenko$^1$, S.V.Fialkovsky$^4$, O.N.Gaponenko$^2$, 
A.A.Garus$^1$, T.I.Gress$^2$, A.M.Klabukov$^1$, A.I.Klimov$^6$, 
S.I.Klimushin$^1$, A.P.Koshechkin$^1$, V.F.Kulepov$^4$, L.A.Kuzmichev$^3$, 
S.V.Lovzov$^2$, B.K.Lubsandorzhiev$^1$, M.B.Milenin$^4$, R.R.Mirgazov$^2$, 
A.V.Moroz$^2$, N.I.Moseiko$^3$, S.A.Nikiforov$^2$, E.A.Osipova$^3$, 
A.I.Panfilov$^1$, Yu.V.Parfenov$^2$, A.A.Pavlov$^2$, D.P.Petukhov$^1$, 
P.G.Pokhil$^1$, P.A.Pokolev$^2$, E.G.Popova$^3$, M.I.Rozanov$^5$, 
V.Yu.Rubzov$^2$, I.A.Sokalski$^1$, Ch.Spiering$^8$, O.Streicher$^8$, 
B.A.Tarashansky$^2$, T.Thon$^8$, R.Wischnewski$^8$, I.V.Yashin$^3$
\end{center}

\begin{center} 
{\it 1 - Institute  for  Nuclear  Research,  Russian  Academy  of   Sciences
(Moscow); \mbox{2 - Irkutsk} State University (Irkutsk); \mbox{3 - Moscow}
State University (Moscow); \mbox{4 - Nizhni}  Novgorod  State  Technical
University  (Nizhni   Novgorod); 5 - St.Petersburg State  Marine
Technical  University (St. Petersburg); \mbox{6 - Kurchatov} Institute
(Moscow); \mbox{7 - Joint} Institute for Nuclear Research (Dubna);
\mbox{8 - DESY} Institute for High Energy Physics (Zeuthen) }
\end{center}

\begin{center}
{\large Presented by {\it I.Sokalski}}
\end{center}

{\bf Abstract}

We review the present status of the Baikal Deep Underwater Neutrino 
Experiment. The construction and performance of the large deep underwater 
Cherenkov detector for muons and neutrinos, {\it NT-200} (Neutrino Telescope 
with 200 phototubes), which is currently under construction in Lake Baikal are
described. Some results obtained with the intermediate detectors {\it NT-36} 
(1993-95), {\it NT-72} (1995-96) and {\it NT-96} (1996-97) are presented, 
including the first clear neutrino candidates selected with 1994 and 1996 data.

\section{Introduction}

The possibility to build a neutrino telescope in Lake Baikal was investigated 
since 1980, with the basic idea to use -- instead of a ship -- the winter ice 
cover as a platform for assembly and deployment of instruments~\cite{chud}. 
After first small size tests, in 1984-90 single-{\it string} arrays equipped 
with 12 -- 36 PMTs ({\it FEU-49} with flat 15 cm photocathode) were deployed 
and operated via a shore cable~\cite{girl}. The total life time for these 
"first generation detectors" made up 270 days. On the methodical side, 
underwater and ice technologies were developed, optical properties of the 
Baikal water as well as the long-term variations of the  water luminescence 
were investigated in great details. For the Baikal telescope site the 
absorbtion length for wavelength between 
470 and \mbox{500 nm} is about 20 m (fig.1), 
%
%
typical value for scattering length is 15 m~\footnote{
               Sometimes the {\it effective scattering length}
               $L_{eff} = L_{scatt} / (1 - \langle \cos \theta \rangle)$ is
               used to characterize the relative merits of different sites for
               neutrino telescopes. With $L_{scatt}$= 15\,m  and 
               $\langle \cos \theta \rangle$ = 0.95  one
               obtains \mbox{$L_{eff}$ = 300\,m} for the Baikal site.
              }
with mean cosine of the scattering angle being close to 0.95 (see~\cite{APP} 
and refs. therein). 
Since 1987, a "second generation detector" with the 
capability to identify muons from neutrino interactions was envisaged. 
Tailored to the needs of the Baikal experiment, a large area hybrid phototube 
{\it QUASAR}~\cite{Pylos} with a hemispherical photocatode of 37 cm diameter 
and a time resolution of better than 3\,nsec was developed to replace the
{\it FEU-49}. According to the approximate number of PMTs this detector was 
named {\it NT-200} -- Neutrino Telescope with 200 PMTs~\cite{baikal}. With an 
estimated effective area of about 2300\,m$^2$ and 8500\,m$^2$ for 1-TeV and 
100-Tev muons, respectively, it is a first stage of a future full-scale 
telescope, which will be built stepwise, via intermediate detectors of rising 
size and complexity. 

\section{The Baikal Neutrino Telescope {\it NT-200}}

The Baikal Neutrino Telescope is being deployed in Lake Baikal, 
\mbox{3.6 km} from shore at a depth of \mbox{1.1 km} (fig. 2,3). 
%
%
It will consist of 192 
optical modules (OMs). The umbrella-like frame carries the 8 strings with the 
detector components. Three underwater electrical cables connect the detector
with the shore station. Deployment of all detector components is carried out 
during 5--7 weeks in late winter when the lake is covered by thick ice.

In April 1993, the first part of {\it NT-200}, the detector {\it NT-36} with 
36 OMs at 3 short strings, was put into operation and took data up to March 
1995. A 72-OM array, {\it \mbox{NT-72}}, run in \mbox{1995-96}. In 1996 it 
was replaced by the four-string array {\it NT-96}. Summed over 700 days 
effective life time, $3.2\cdot 10^{8}$ muon events have been collected with 
\mbox{{\it NT-36, -72, -96}}. Since \mbox{April 6,} 1997, {\it NT-144}, a 
six-string array with 144 OMs, is taking data in Lake Baikal.

The OMs are grouped in pairs along the strings. The pulses from two PMTs of a 
pair after \mbox{0.3 {\it p.e.}} discrimination are fed to a coincidence with 
\mbox{15 ns} time window. A pair defines a {\it channel}. A {\it muon-trigger} 
is formed by the requirement of \mbox{$\geq N$ {\it hits}} (with {\it hit} 
refering to a channel) within \mbox{500 ns}. $N$ is typically set to the value 
\mbox{3 or 4.} For  such  events, amplitude and time of all fired channels are
digitized and sent to shore. The event record includes all hits within a time 
window of -1.0 $\mu$sec to +0.8 $\mu$sec with respect to the muon trigger 
signal. A separate {\em monopole trigger} system searches for time patterns 
characteristic for slowly moving objects (see Subsect.4.4).

Stability of digitizing is checked by calibration runs performed typically 
once per several days. The calibration of the relative time shifts between all
channels is performed with the help of a nitrogen laser with 300 ps pulse 
width positioned above the array. The light from this laser is guided by 
optical fibers of equal length separately to each OM pair. To cross check this
method, for {\it NT-144} a special second laser emitting light directly 
through the water was mounted at one string 20m below the last layer of OMs. 
In order to monitor the performance of the OMs, the counting rates of 
individual PMTs as well as of the channels are measured and transmitted to 
shore station once per 60 sec. To investigate short time variations of 
counting rates, in {\it NT-144} a new scaler system has been installed which 
measures the counting rates of all channels with 0.8 sec step.

In the initial project of {\it NT-200}, the optical modules are grouped in 
pairs along the strings directed alternatively upward and downward. The 
distance between pairs looking face to face is 7.5 m, while pairs arranged 
back to back are 5 m apart. In this case the array has a symmetrical response 
to upward and downward muons, respectively. We tested this orientation of OMs 
with {\it NT-36} and {\it NT-72}. However, due to sedimentation the 
sensitivity of uplooking OMs decreased by 50\% after 150 days. Hence for 
{\it NT-96} and {\it NT-144} the orientation of OMs has been changed: only OMs
from two layers of the array (the second and eleventh) look upward, and all 
others look downward. Nevertheless we possibly come back toward symmetrical 
structure if the problems with sedimentation will be overcame.

\section{Track Reconstruction}

In contrast with a typical underground detector, it is impossible to determine
co-ordinates for some clearly visible points which would belong to a track of 
a particle crossing an underwater array because it represents a lattice of OMs
with large distances between them. The parameters of a muon 
track~\cite{APP,Reco} crossing an underwater detector have to be determined by 
minimizing~\footnote{ 
       The reconstruction algorithm is based on the assumption that the
       light radiated by the muons is emitted exactly under the  Cherenkov
       angle (42 degrees) with respect to the muon path. This "naked muon 
       model" is a simplification, since the direction of shower particles 
       accompanying the muons is smeared around the muon direction.} 

\begin{equation}
\chi^2_t = \sum_{i=1}^{N_{hit}} (T_i(\theta, \phi, u_0, v_0, t_0)
    - t_i)^2 / \sigma_{ti}^2
\end{equation}

\noindent
Here, $t_i$ are the measured times and $T_i$ the times expected for a given 
set of track parameters. $N_{hit}$ is the number of hit channels, 
$\sigma_{ti}$ are the timing errors. A set of parameters defining a straight 
track is given by $\theta$ and $\phi$ -- zenith and azimuth angles of the 
track, respectively, $u_0$ and $v_0$ -- the two coordinates of the track point 
closest to the center of the detector, and $t_0$ -- the time the muon passes 
this point. For the results given here we do not include an amplitude term 
$\chi^2_a$ analog to $\chi^2_t$ in the analysis, but use the amplitude 
information 
only to calculate the timing errors  $\sigma_{ti}$ in the denominator of the 
formula above. Only events fulfilling the condition  \mbox{``$\geq 6$} hits at
\mbox{$\geq 3$} strings`` are selected for the standard track reconstruction 
procedure which consists of the following steps:

\noindent
{\bf 1.} A preliminary analysis includes several causality criteria rejecting 
events which violate the model of a naked muon. After that, a 0-th 
approximation of $\theta$ and $\phi$ is performed.

\noindent
{\bf 2.} The $\chi^2$ minimum search (minimization of the function eq.1), 
based on the model of a naked muon and using only time data.

\noindent
{\bf 3.} Quality criteria to reject most badly reconstructed events.

We have developed a large set of pre-criteria as well as quality criteria to 
reject misreconstructed events. Most of these criteria are not independent of 
each other.  Furthermore, the optimum set of criteria turned out to depend on 
the detector configuration. The causality criteria refer to time differences  
between channels. {\it E.g.}, one requests that each combination of two 
channels $i,j$ obeys the condition $c~|dt_{ij}| < n~|dx_{ij}| + c~\delta t$,
where $dt_{ij}$ and $dx_{ij}$ are time differences and distances between 
channels $i$ and $j$, respectively and $n = 1.33$ is refraction coefficient 
for water. The term $\delta t =$ 5\,nsec accounts for the time jitter. Some of
the most effective quality criteria are, {\it e.g.}, upper limits on 
parameters like the minimum $\chi^2$, the probability $P_{nohit}$ of non-fired 
channels not to respond to a naked muon and probability $P_{hit}$ of fired 
channels to respond to a naked muon. 

\section{Selected Results}

\subsection{Atmospheric Muons Vertical Flux}

Muon angular distributions are well described by MC expectations. Converting 
the measured angular dependence obtained with the standard reconstruction 
procedure (see Sect.3) applied to {\it NT-36} data~\cite{APP} into a depth 
dependence of the vertical flux, good agreement with theoretical 
predictions~\cite{bug} as well as with others experimental 
points~\cite{vertical} is observed (see fig.4).
%
%

\subsection{Separation of Neutrino Events with Standard Track Reconstruction}

The most obvious way to select events from the lower hemisphere (which 
dominantly are due to atmospheric neutrino interactions in the ground or water
below the array) is to perform the full spatial track reconstruction (see 
Sect.3) and select the events with negative $\theta$ values. Taking into 
account that the flux of downward muons is about 6 orders of magnitude larger 
than the flux of upward muons, the reconstruction procedure should be 
performed extremely thoroughly. Even if very small fraction of downward 
atmospheric muons is misreconstructed as up-going ones, it forms an essential 
background. Due to small value of $S/N$ ratio (where $S$ is counting rate of 
upward neutrino induced events and $N$ is counting rate of downward 
atmospheric muons which are reconstructed as upward events), it is impossible 
to observe clear neutrino signal with {\it NT-36} and {\it NT-72} data and the
current level of standard reconstruction procedure. MC calculations indicate 
the essetially better characteristics for {\it NT-96} detector which can be 
considered as a neutrino telescope in the wide region of $\theta$. 

The analysis of the {\it NT-96} data with respect to neutrino induced upward 
muons using the standard reconstruction procedure is presently in 
progress~\cite{neutrino97}. We apply final quality cuts after the minimization 
(see Sect.3). For {\it NT-96} the most effective cuts are the traditional 
$\chi^2$ cut, cuts on the probability of non-fired channels not to be hit, and 
fired channels to be hit ($P_{nohit}$ and $P_{hit}$, respectively), cuts on 
the correlation function of measured amplitudes to the amplitudes expected for
the reconstructed tracks, and a cut on the amplitude $\chi^2$ defined similar 
to the time $\chi^2$ defined above. To guarantee a minimum lever arm for track 
fitting, we reject events with a projection of the most distant channels on 
the track ($Z_{dist}$) below 35 meters. Due to the small transversal 
dimensions of {\it NT-96}, this cut excludes zenith angles close to the 
horizon, i.e., the effective area of the detector with respect to atmospheric 
neutrinos is decreased considerably (fig.5).
%
%

The efficiency of all criteria has been tested using MC generated atmospheric
muons and upward muons due to atmospheric neutrinos. $ 1.8 \cdot 10^6$ events 
from atmospheric muon events (trigger {\it 6/3}) have been simulated, with
only 2 of them passing all cuts and being reconstructed as upward going muons.
This corresponds to $S/N \approx 1$. Rejecting all events with less than 9 
hits, no MC fake event is left, with only a small decrease in neutrino 
sensitivity. This corresponds to $S/N > 1$ and the lowest curve in fig.5.

The table shows the fraction of events after the final quality criteria, 
normalized to the number of events surviving pre-criteria and reconstruction, 
for triggers {\it 6/3} and {\it 9/3}, respectively:

\begin{center}
\begin{tabular} {|c|c|c|c|} \hline
\raisebox{0pt}{Trigger cond.}&
\raisebox{0pt}{Experiment}&
\raisebox{0pt}{MC atm $\mu$}&
\raisebox{0pt}{MC $\mu$ from $\nu$} \\ \hline \hline
6/3 &   0.19 &           0.21    &        0.20 \\ \hline
9/3 &   0.044&           0.056   &       0.175 \\ \hline
\end{tabular}
\end{center}

With this procedure, we have reconstructed $5.3 \cdot 10^6$ events taken with 
{\it NT-96} in April/May 1996. The resulting angular distribution is presented 
in fig.6. 
%
%
Three events were recognized as upward going muons. Fig.7 displays one of the 
neutrino candidates. 
%
%
Top right the times of the hit channels are shown as as function of the 
vertical position of the channel. At each string we observe the time 
dependence characteristically for upward moving particles.

One of the causality criteria (see item 1 in Sect.3) demands that the zenith
angle regions ${\theta^{min} -\theta^{max}}$ consistent with the observed time 
differences $\Delta t_{ij}$  between two channels {\it i}, {\it j} along the
same string

\begin{equation}
\label{eq:thetalimit}
\cos(\theta^{min}+\eta) < \cos\theta \frac{c \cdot \Delta t_{ij}}{z_j-z_i} < \cos({\theta^{max}-\eta})
\end{equation}

\noindent
(here ${z_i,z_j}$ are z coordinates of the channels and $\eta$ is the Cherenkov
angle) must overlap for all channels along the string. Applying 
eq.\ref{eq:thetalimit} not only to pairs at the same string, but to all pairs 
of hit channels, one can construct an allowed region in both $\theta$ and 
$\phi$. For clear neutrino events this region lays totally below horizon. 
This is demonstrated at the bottom right picture of fig.7. The same holds for 
the other two events, one of which is shown in fig.8a. Fig.8b, in contrast, 
shows an ambiguous event giving, apart from the upward solution, also a 
downward solution. In this case we assign the event to the downward sample.
%
%

The analysis presented here is based on the data taken with {\it NT-96} 
between April 16 and May 17, 1996 (18 days lifetime). Three neutrino 
candidates have been separated, in good agreement with the expected number of 
upward events of approximately 2.3. Our algorithm allows to select neutrino 
events in a cone with about 50 degrees half-aperture around the opposite 
zenith, and an effective area of $\sim 350 m^2$. 

We hope that the further analysis of {\it NT-96} and {\it NT-144} data will 
confirm our capability to select the neutrino induced events over the 
background of fake events from downward atmospheric muons.  With the 
experimental confirmation that {\it NT-96} can operate as a neutrino detector,
we now are searching for additional possibilities to reject fake events with a
smaller loss in effective area. The increased transversal dimensions of the 
future {\it NT-200} (1998) will significantly increase effective area and 
angular acceptance for reliably separable up-going events.

\subsection{Search for Nearly Upward Moving Neutrinos}

To identify nearly vertically upward muons with energies below 1 TeV (as 
expected, {\it e.g.}, for muons generated by neutrinos resulting from 
neutralino annihilation in the core of the Earth), full reconstruction is 
found to be not neccessary~\cite{ourneu}. Instead, separation criteria can be 
applied which make use of two facts: firstly, that the muons searched for have
the same  vertical direction like the string; secondly, that low-energy muons 
generate mainly direct Cherenkov light and, consequently, are not visible over
large distances and should produce a clear time and amplitude pattern in the 
detector. We have choosen the following separation criteria: 

\noindent
{\bf 1.} Time differences between any two hit channels $i$ and $j$ must obey 
the inequality

\begin{equation}
             |(t_{i}-t_{j})-(T_{i}-T_{j})|<dt
\end{equation}

\noindent
where $t_i(t_j)$ are the measured times in channels $i(j)$, $T_i(T_j)$ are 
the ``theoretical'' times expected for minimal ionizing, up-going vertical 
muons and $dt$ is a time cut. 

\noindent

{\bf 2.} The minimum value of amplitude asymmetries for all pairs of 
alternatively directed hit channels must obey the inequality

\begin{equation}        
dA_{ij}(down-up) > 0.3, 
\end{equation}

\noindent
where $dA_{ij}(down-up)=(A_{i}(down)-A_{j}(up))/(A_{i}(down)+A_{j}(up))$ and 
$A_{i}(down) (A_{j}(up))$ are the amplitudes of channel $i(j)$ facing 
downward(upward).

\noindent
{\bf 3.} All amplitudes of downward looking hit channels must exceed 4 
photoelectrons:

\begin{equation}        
A_i(down)>4 ph.el.
\end{equation}

\noindent
{\bf 4.} The amplitude asymmetry   $dA(down-down)$ for downward looking hit 
channels is defined as that of the 3 possible combinations 
$dA_{ij}(down-down)=(A_i-A_j)/(A_i+A_j)\mid _{i>j}$ with the largest absolute 
value. For background events due to showers below the array it peaks at values 
close to 1, for vertical neutrino candidates it should be close to zero. The 
fourth criterion rejects half of the neutrino sample and nearly all events due 
to deep showers from downward atmospheric muons:

\begin{equation}        
            dA(down-down)<0.
\end{equation}

The dependence of the expected yearly number of muons generated by atmospheric 
neutrinos and of background events on the time cut $dt$  are presented in 
fig.9.
%
%
The curves marked 1, 2, 3 and 4 correspond to the trigger conditions {\it 1, 
1-2, 1-3} and {\it 1-4}, respectively. The 'crosses' denote background curves 
and 'asterisks' denote muons from atmospheric neutrinos. One sees that the 
signal-to-noise ratio $S/N$ is close to 1 for trigger conditions {\it 1-3} and 
$dt\leq20ns$ and improves with decreasing $dt$ or applying criterion {\it4}.

The  analysis presented here is based on the data taken with {\it \mbox{NT-36}}
between April 8, 1994 and March 5, 1995 (212 days lifetime). There were 6 PMT
pairs along each of the 3 strings of NT-36. The orientation of the channels 
from top (channel \#1) to bottom (channel \#6) at each string was 
{\em up-down-up-down-up-down} in 1994/95. Upward-going muon candidates were 
selected from a total of \mbox{$8.33\cdot 10^{7}$} events recorded by the 
muon-trigger \mbox{"$\geq 3$} hit channels". The samples fulfilling trigger 
conditions {\it 1, 1-2, 1-3} and {\it 1--4} with time cut $dt=20$ns contain 
131, 17 and 2 events, respectively. Only two events fulfill trigger conditions 
{\it 1--3} and {\it 1--4}. These events were recorded at 6 June and 3 July 
1994. The first event is consistent with a nearly vertical upward going muon 
and the second one with an upward going muon with zenith angle
$\theta_{\mu}=15^{\circ}$ (fig.10).
%
%

Fig.11 shows the passing rate for two samples of events in dependence on the
time cut $dt$. 
%
%
The "experimental neutrino sample" consists of just the two events shown in 
fig.10, the "experimental background sample" contains all other events with 
the exception of these two. MC curves have been obtained from modelling upward 
muons from atmospheric neutrinos ("neutrinos") and from downward going
atmospheric muons ("background").

Fig.11a demonstrates that MC describes the data within a factor of 3-4. From 
fig.11b one sees that the probability to observe a background event with 
$dt<20$nsec is about 2 percent only. Whereas the shapes of  experimental and 
MC  distributions in fig.11b are quite similar, the absolute values disagree by
a factor of 1.5-4, depending on the criterion. The MC calculated numbers of 
upward going muons are systematically below the two experimentally observed 
events. Apart from statistics, the reason may be the following: MC simulations
of the NT-36 response to upward going muons from atmospheric neutrinos has 
been performed without taking into account light scattering in water. A raw 
estimate shows that the expected number of detected upward going muons may 
rise by 40-80\% when scattering process will be taken into account.  

Considering the two neutrino candidates as atmospheric neutrino events, a 
90 \% CL upper limit of $1.3 \cdot 10^{-13}$ (muons cm$^{-2}$ sec$^{-1}$) in 
a cone with 15 degree half-aperture around the opposite zenith is obtained for
upward going muons generated by neutrinos due to neutralino annihilation in the
center of the Earth. The limit corresponds to muons with energies greater than
the threshold energy $E_{th} \approx 6$ GeV, defined by 30m string length. 
This is still an order of magnitude higher than the limits obtained by 
Kamiokande~\cite{kamneutralino}, Baksan~\cite{bakneutralino} and 
MACRO~\cite{MACROneutralino}. The effective area of {\it NT-36} for nearly 
vertical upward going muons fulfilling our separation criteria {\it 1-3} with 
$dt=20ns$ is $S_{eff}=50$ m$^{2}$/string. A rough estimate of the effective 
area of the full-scale Baikal Neutrino Telescope {\it \mbox{NT-200}} (with 
eight strings twice as long as those of {\it \mbox{NT-36}}) with respect to 
nearly vertically upward going muons gives $S_{eff} \approx 400-800$ m$^{2}$.

\newpage

\subsection{Search for Magnetic Monopoles}

\subsubsection{Monopoles Catalyzing Baryon Decay}

It was shown by Rubakov~\cite{ruba} and Callan~\cite{callan} that baryon 
number violation is possible in the presence of a GUT magnetic monopole. For 
reasonable velocities  ($\beta \leq 10^{-3}$), a catalysis cross section

\begin{equation}
\sigma_c = 0.17 \cdot \sigma_o / \beta^{2}
\end{equation}

is predicted for monopole-proton interactions~\cite{araf} with $\sigma_o$ 
being of the order of magnitude typical for strong interactions. Following 
these predictions, the average distances between two sequential proton decays  
along the monopole track in water can be as short as 10$^{-2}$ - 10$^{1}$ cm.

In order to search for GUT monopoles, a dedicated trigger system was 
implemented in the electronics scheme. It is based on the method which has 
been developed for the first-stage Baikal setups GIRLYANDA~\cite{girl} which 
operated in 1984-89. The method selects events which are defined as a 
short-time (0.1 - 1 msec) \mbox{1 {\it p.e.}} increase of the counting rate of
individual channels. This pattern is expected from sequential Cherenkov 
flashes produced by a monopole along it's track via the proton decay products.
Due to the large $\sigma_c$ values, for a short time interval the rate of 
detected flashes can appreciably exceed the counting rate from PMT dark 
current noise and  water luminescence, even for monopoles passing a channel at
a distance of several tens of meters.

Our monopole system consists of several nearly independent modules. They are 
synchronized by a common 10 kHz clock. One module reads the signals of 6
neighbouring channels placed on the same string. During standard data taking 
runs, the monopole trigger condition was defined as $\geq3$ hits  within
a time window of 500 $\mu$sec in  any  of  the  channels (values for number 
of hits and time windows  duration  can be set from shore). Once the trigger
condition is fulfilled, the information about the number of hits in each of 
the 6 channels which occur within the given time window is sent to shore. 
Time and amplitude information is not recorded by the monopole system.

Since the method is based on the search for counting rate excesses, it was 
important to perform a long-term {\it in situ} check and to verify that the 
time behaviour of local triggers is described by a Poisson  distribution. 
If not, the search for monopoles via counting rate splashes would become 
essentially complicated. It was elucidated that non-Poisson effects are 
suppressed effectively by the coincidence between the two PMTs of a pair. 
This can be seen from fig.12. 
%
%
We compare the number of hits detected within 8 msec-time windows to the 
calculated distribution. Calculating it, we assumed pure Poisson fluctuations 
around the independently measured average hit rate. Such good agreement with 
the Poisson assumption is observed for all channels with only very few 
exceptions.

The data taken from April 16th to November 15th 1993 with {\it NT-36} have been
analyzed with respect to monopoles  catalyzing baryon decay. Using the 
off-line threshold \mbox{$\geq7$ hits} within \mbox{500 $\mu$sec} we rejected
most events, registrating only the uppermost tail of the Poisson distribution.
To suppress the remaining accidental noise, we defined an even tighter 
trigger, requesting that one channel had counted \mbox{$\geq7$ hits} and the 
second channel looking to it's face and situated \mbox{7.5 m} away along the
same string had counted \mbox{$\geq3$ hits} in the same \mbox{500 $\mu$sec} 
(two channels looking face to face define a {\it svjaska} which is one of the 
main levels in the hierarchy of the Baikal array~\cite{baikal}. This reduces 
the number of the experimentally observed monopole candidates to zero. Fig.13 
shows the number of hits from the \mbox{channel 1} of a svjaska plotted 
versus the number of hits  from  the face-to-face \mbox{channel 2.} 
%
%
The product of effective  data  taking  time  and number of \mbox{svjaskas} 
with both channels operating is 4573 hours for the investigated period.
\mbox{$3.5\cdot 10^{7}$} monopole triggers with \mbox{$\geq3$ hits} have been 
taken.

After calculating the effective area, from the non-observation of monopole 
canditates we obtain the upper flux limits (90 $\%$ CL) shown in  fig.14  
together with our earlier results~\cite{girl}, results from IMB~\cite{IMBmon}
and KAMIOKANDE~\cite{kammon} and with the theoretical Chudakov-Parker 
limit~\cite{CP}. A limit of $2.7\cdot 10^{-16}$ cm$^{-2}$ s$^{-1}$ has been 
obtained by the Baksan Telescope~\cite{bakmon} for $\beta > 2 \cdot 10^{-4}$.
%
%

The present analysis is rather straight-forward using the same tight trigger 
condition for all time periods. This trigger suppresses fake monopole 
candidates even during periods with a high level of water luminescence. It is 
planned to tune the trigger depending on the local trigger counting rate. Using
this more sophisticated trigger and the whole statistics taken for 4.5 years 
with {\it NT-36, -72, -96} and {\it -144}  (more than 800 day's data sample by 
September 1997) one can decrease the minimal detectable fluxes by a factor 
10 -- 100 compared to the results presented here. A further considerable 
progress is expected with the whole {\it NT-200}.

\subsubsection{Search for Fast Magnetic Monopoles}

The basic mechanism for light generation by a relativistic  magnetic monopole
is Cherenkov radiation. The large magnetic charge results in a giant light 
intensity, equal to that  of  a  14-PeV  muon  for a relativistic monopole 
with  magnetic  charge \mbox{{$\it g_o = 68.5e$}.} One  can search for such
monopoles analysing the data obtained with the muon trigger. Due to the 
non-stochastical nature of the Cherenkov light emission of relativistic 
monopoles (contrary to a 14 PeV muon!), there is a close correspondence 
between reconstructed monopole track parameters and the number of hit 
channels. This can be effectively used  to  select monopole candidates. For 
{\it NT-200}, we estimate an effective area roughly as a of 
{$\approx 2\cdot 10^{4}$} m$^2$ with respect to monopoles with 
$\beta \approx 1$. More detailed calculations which take into account the 
background from fake events produced by atmospheric muons are needed, but 
preliminary study shows that these events can be rejected rather effectively.  

The data provided by the muon trigger can be used to search for monopoles with
velocities down to $\beta \approx 0.1$. As was mentioned in Sect.2, the 
time and amplitude information is collected for all  channels hit within a 
time window of 1.8 $\mu$sec around the muon trigger. This time is comparable 
with the time needed by particles with velocities $\beta \approx 0.1$ to cross
the array volume. For velocities $0.6\leq \beta \leq 0.75$, most light is due 
to Cherenkov radiation generated by $\delta$-electrons knocked out by the 
monopole. For $\beta \leq 0.6$ the basic mode of light emission is water 
luminescence~\cite{trof}. The luminescence of the Baikal water is extremely 
small (about one photon per 5 MeV energy loss, as we have studied 
experimentally with $\alpha$-particles). Nevertheless, due to the giant energy
released, the luminescence stimulated by a monopole with charge {$\it g_{o}$} 
exceeds the Cherenkov light emitted by a relativistic muon down to monopole 
velocities of $\beta \approx 0.1$. The effective area of {\it NT-200} for 
monopoles with $\beta = 0.5$ and $\beta = 0.2$ is estimated as 1000  m$^2$ and
500 m$^2$, respectively.

We plan to analyse data which taken with the various Baikal arrays with 
respect to the search for fast magnetic monopoles.

\section{The Next Steps}

On April 6, 1997, a six-string array with 144 optical modules, {\it NT-144},
was put into operation. By September 1, 1997 it has collected $\approx 1.5 
\cdot 10^{8}$ muons. We plan to complete the {\it NT-200} array in April,
1998.   

\section{Acknowledgements}
This work was supported by the Russian Ministry of Research,the German 
Ministry of Education and Research and the Russian Fund of Fundamental 
Research ( grants {\sf 96-02-17308} and {\sf 97-02-31010}).

\newpage

\strut

\vspace{3cm}

\begin{figure}[h] 
\centering
\mbox{\epsfig{file=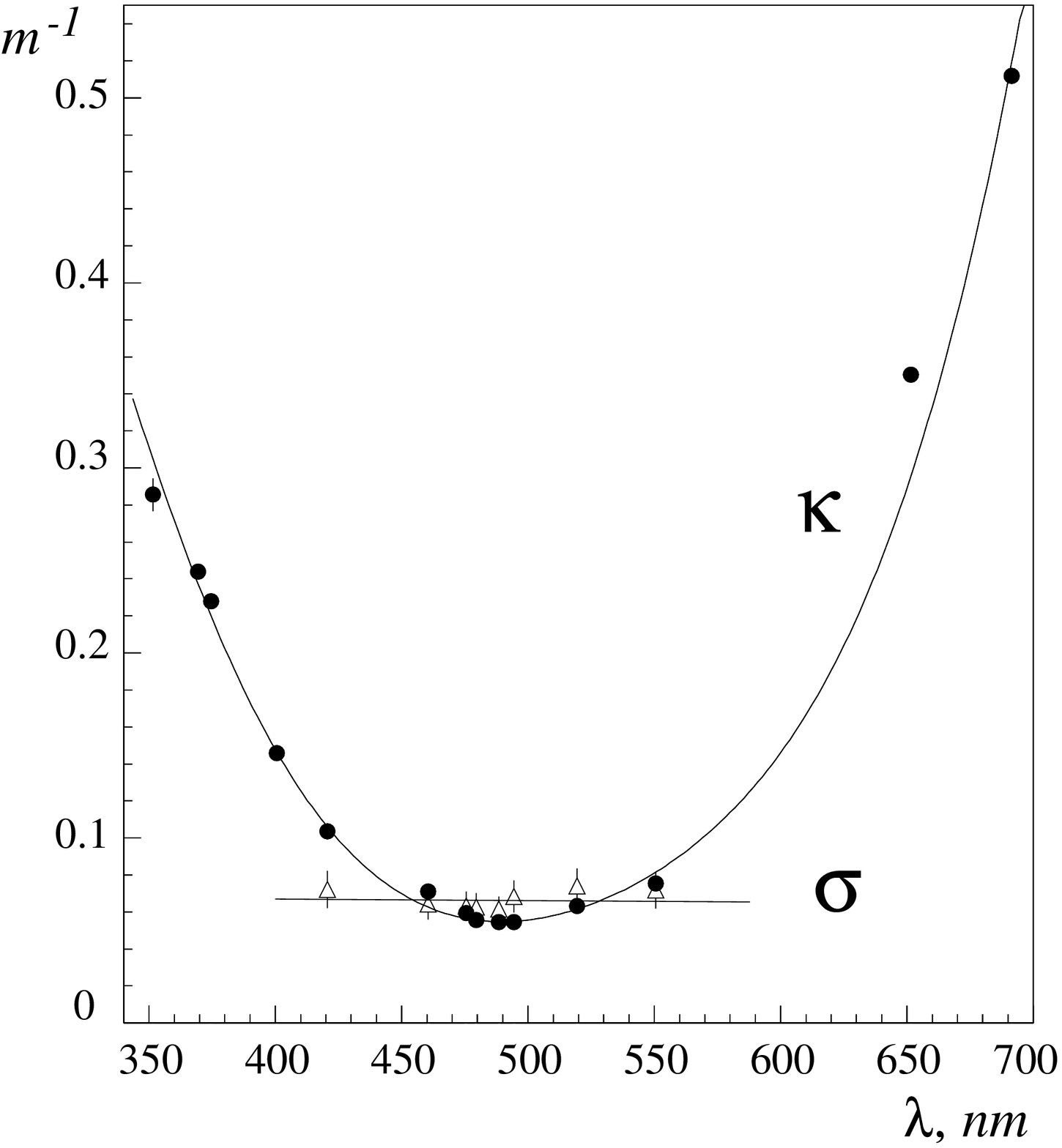,height=14cm}}
\end{figure}

\vspace{1cm}

{\bf Figure 1:} Coefficients for light absorption ($\kappa$) (dots) and 
scattering ($\sigma$) (triangles)  at the site of the Baikal Neutrino 
Telescope as a function of wavelength. Date of measurement:  Oct./Nov.1993. 
Depth: 1100\,m.

\newpage

\strut

\vspace{1cm}

\begin{figure}[h] 
\centering
\mbox{\epsfig{file=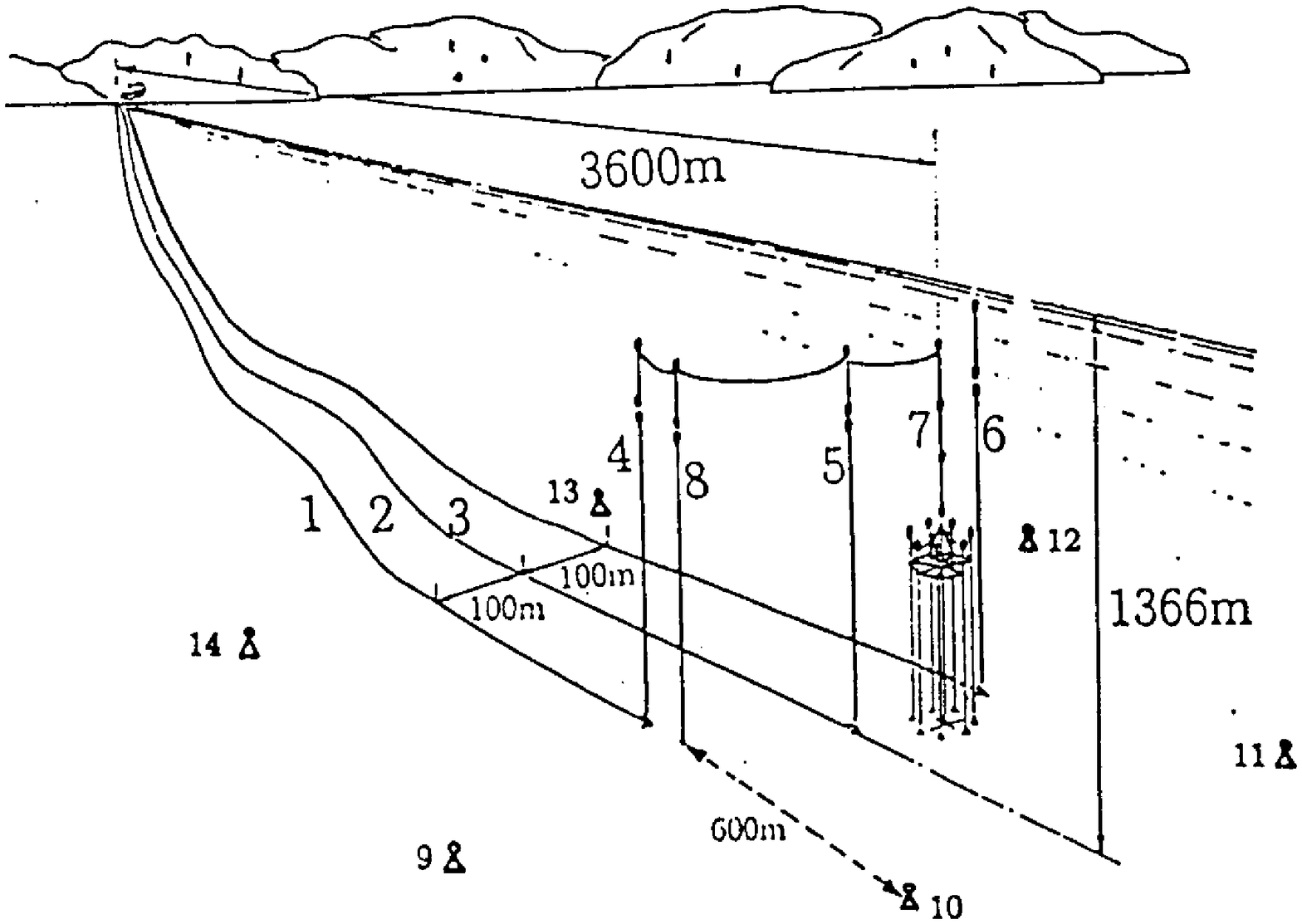,height=12cm,angle=90}}
\end{figure}

\vspace{1cm}

{\bf Figure 2:} The Baikal detector complex (status since 1994). {\it 1,2} -- 
wire cables to shore, {\it 3} -- opto-electrical cable to shore, {\it 4,5,6} 
-- string stations for shore cables 1,2,3, respectively, {\it 7} -- string 
with the telescope, {\it 8} -- hydrometric string, {\it 9-14}  -- ultrasonic 
emitters.

\newpage

\strut

\vspace{0cm}

\hspace{-1cm}
\begin{figure}[h] 
\mbox{\epsfig{file=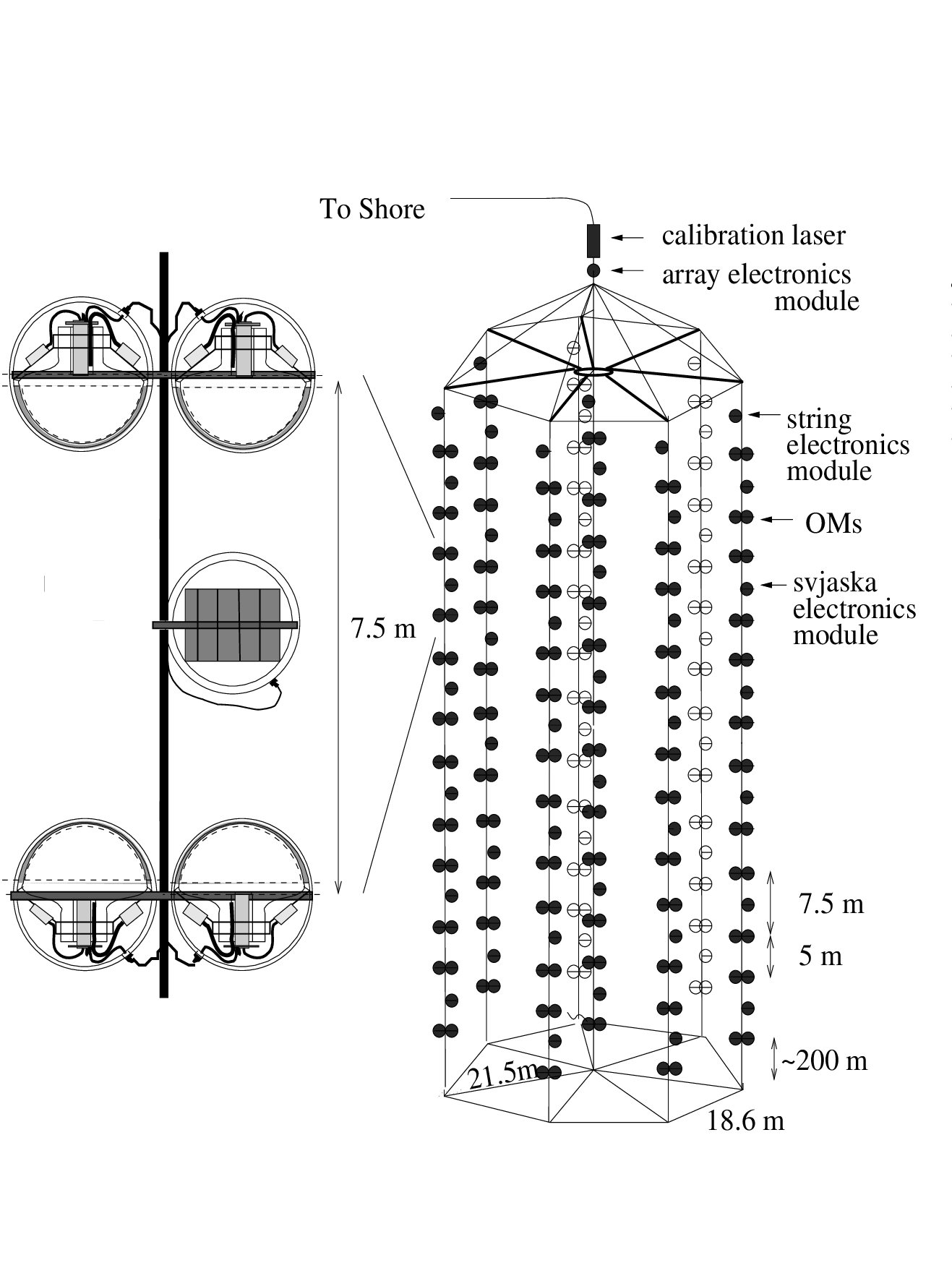,height=16cm,width=11.3cm}}
\end{figure}

\vspace{0cm}

{\bf Figure 3:} Schematic view of the Baikal Telescope {\it NT-200}. The 
modules of {\it NT-144}, operating since April 6. 1997 are in black, The 
expansion left-hand shows 2 pairs of optical modules ("svjaska") with the 
svjaska electronics module, which houses parts of the read-out and control 
electronics. 

\newpage

\strut

\vspace{1cm}

\begin{figure}[h] 
\centering
\mbox{\epsfig{file=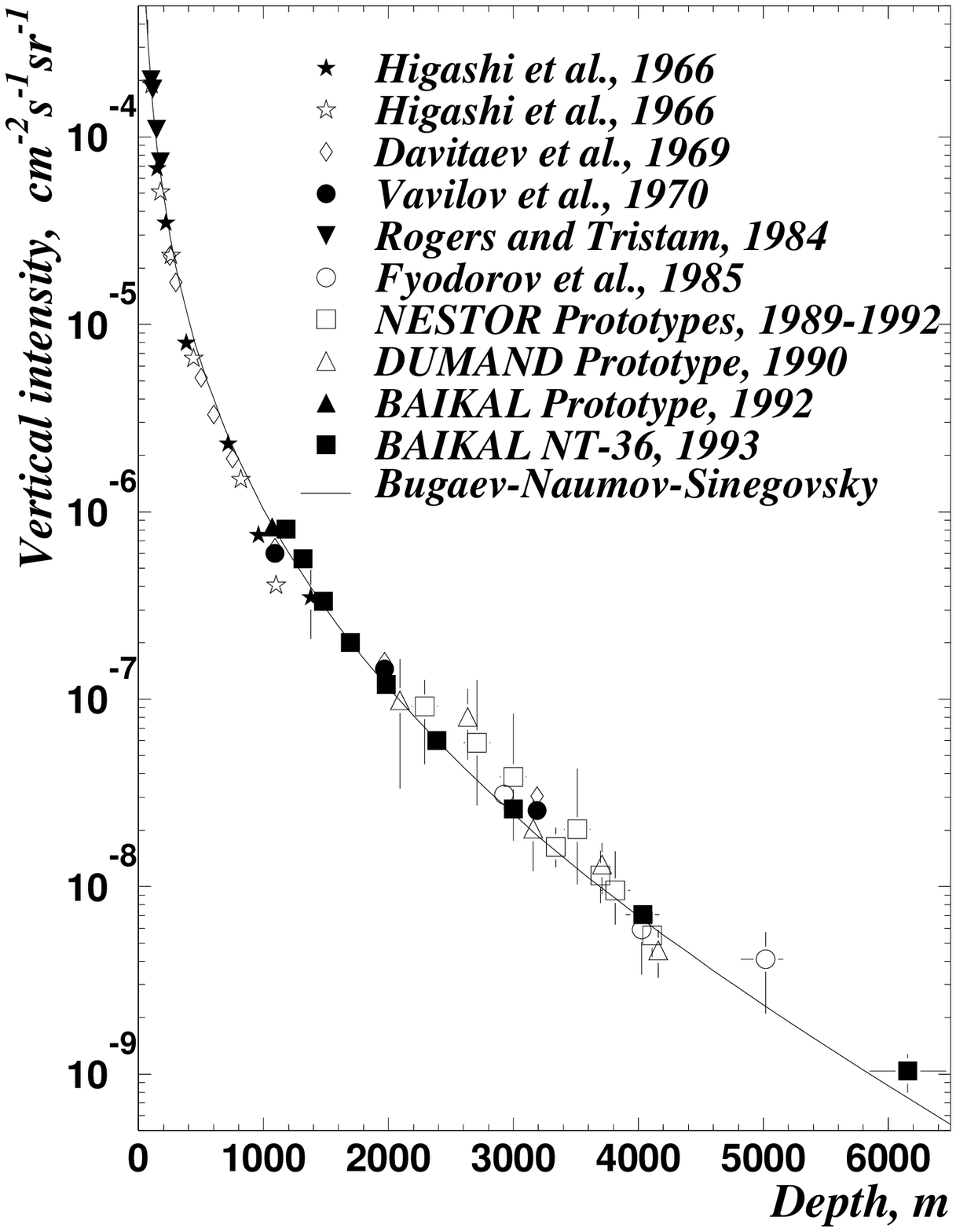,height=15cm}}
\end{figure}

\vspace{1cm}

{\bf Figure 4:} Vertical muon flux, $I_{\mu}(cos\theta=1)$, vs. water depth 
$L$. The nine {\it NT-36} values (full squares) are calculated for 
$\cos\theta$\,=\, 0.2 to 1.0 in steps of 0.1. The other data points are taken 
from refs.~\cite{vertical}. The curve is taken from~ \cite{bug}. 

\newpage

\strut

\vspace{1cm}

\begin{figure}[h] 
\centering
\mbox{\epsfig{file=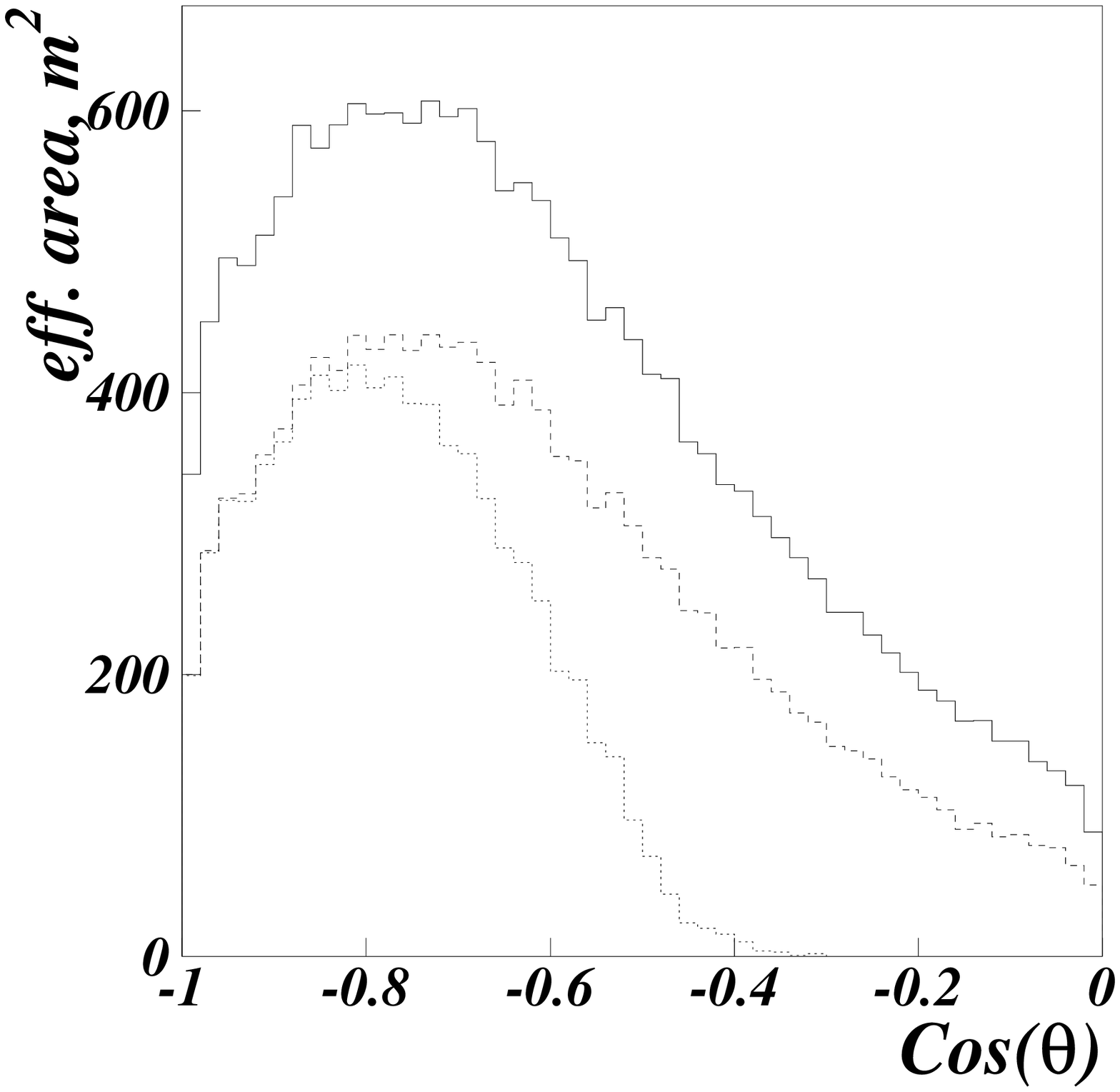,height=15cm}}
\end{figure}

\vspace{1cm}

{\bf Figure 5:} Effective area for upward muons satisfying trigger {\it 9/3};
solid line -- no quality cuts; dashed line -- final quality cuts; dotted line 
-- final quality cuts and restriction on $Z_{dist}$ (see text).

\newpage

\strut

\vspace{1cm}

\begin{figure}[h] 
\centering
\mbox{\epsfig{file=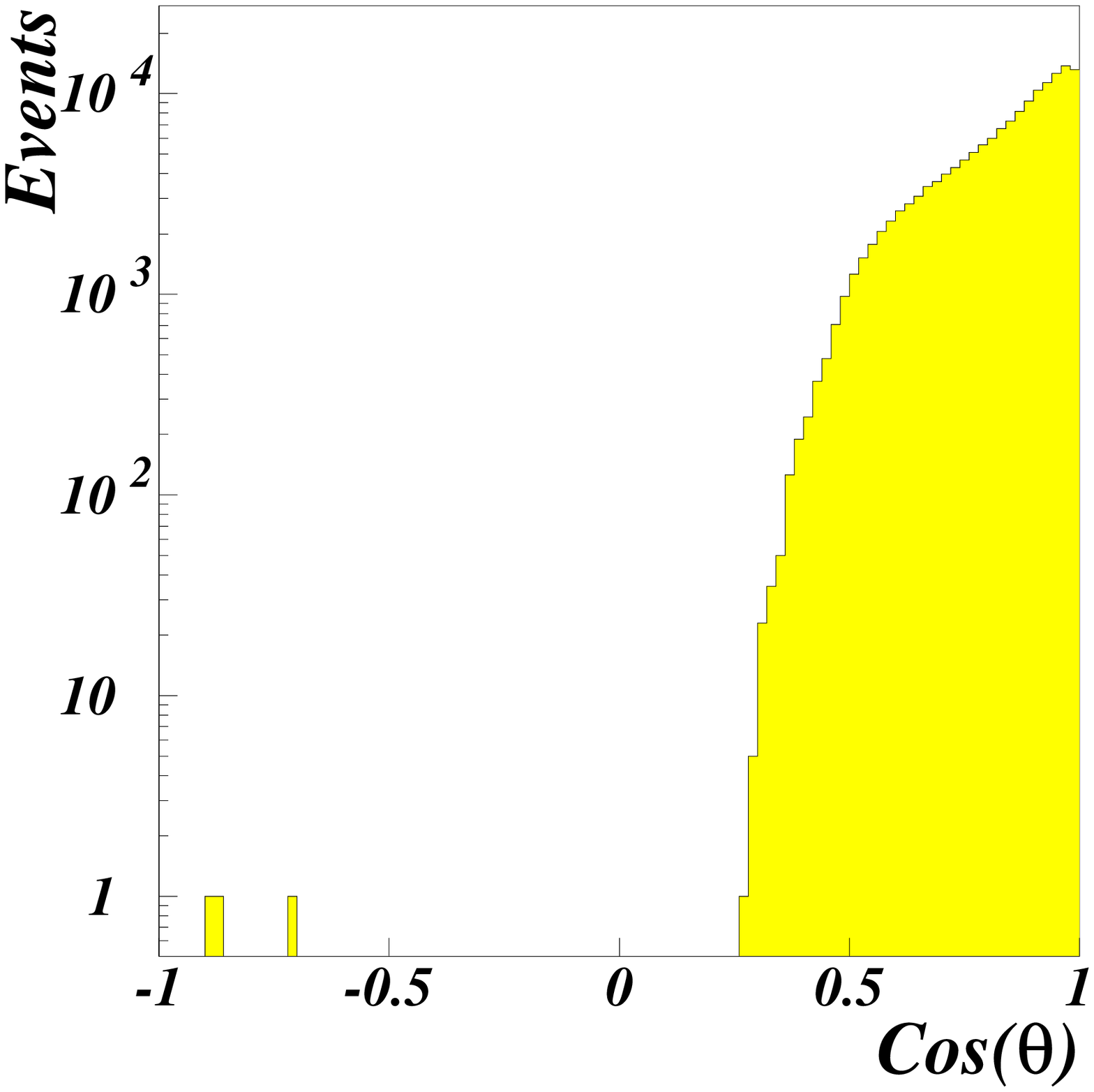,height=15cm}}
\end{figure}

\vspace{1cm}

{\bf Figure 6:} Experimental angular distribution of events satisfying trigger 
{\it 9/3}, all final quality cuts and the limit on $Z_{dist}$ (see text).

\newpage

\strut

\vspace{1cm}

\begin{figure}[h] 
\centering
\mbox{\epsfig{file=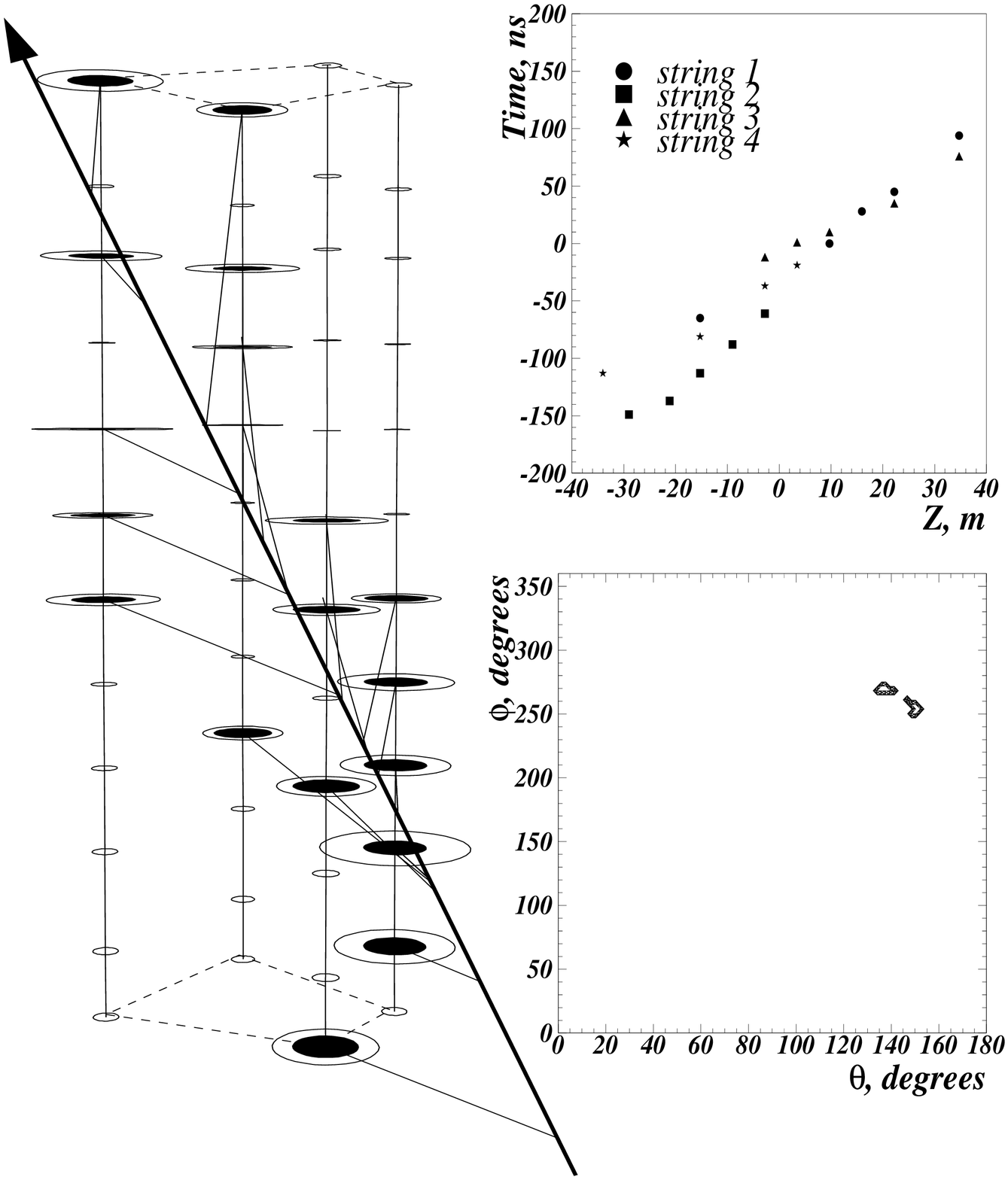,height=16cm}}
\end{figure}

\vspace{0cm}

{\bf Figure 7:} A "gold plated" 19-hit neutrino event. {\it Left:} Event 
display. Hit channels are in black. The thick line gives the reconstructed 
muon path, thin lines pointing to the channels mark the path of the Cherenkov 
photons as given by the fit to the measured times. The areas of the circles 
are proportional to the measured amplitudes. {\it Top right:} Hit times versus
vertical channel positions. {\it Bottom right:}  The allowed $\theta/\phi$ 
regions (see text).

\newpage

\strut

\vspace{1cm}

\begin{figure}[h] 
\centering
\mbox{\epsfig{file=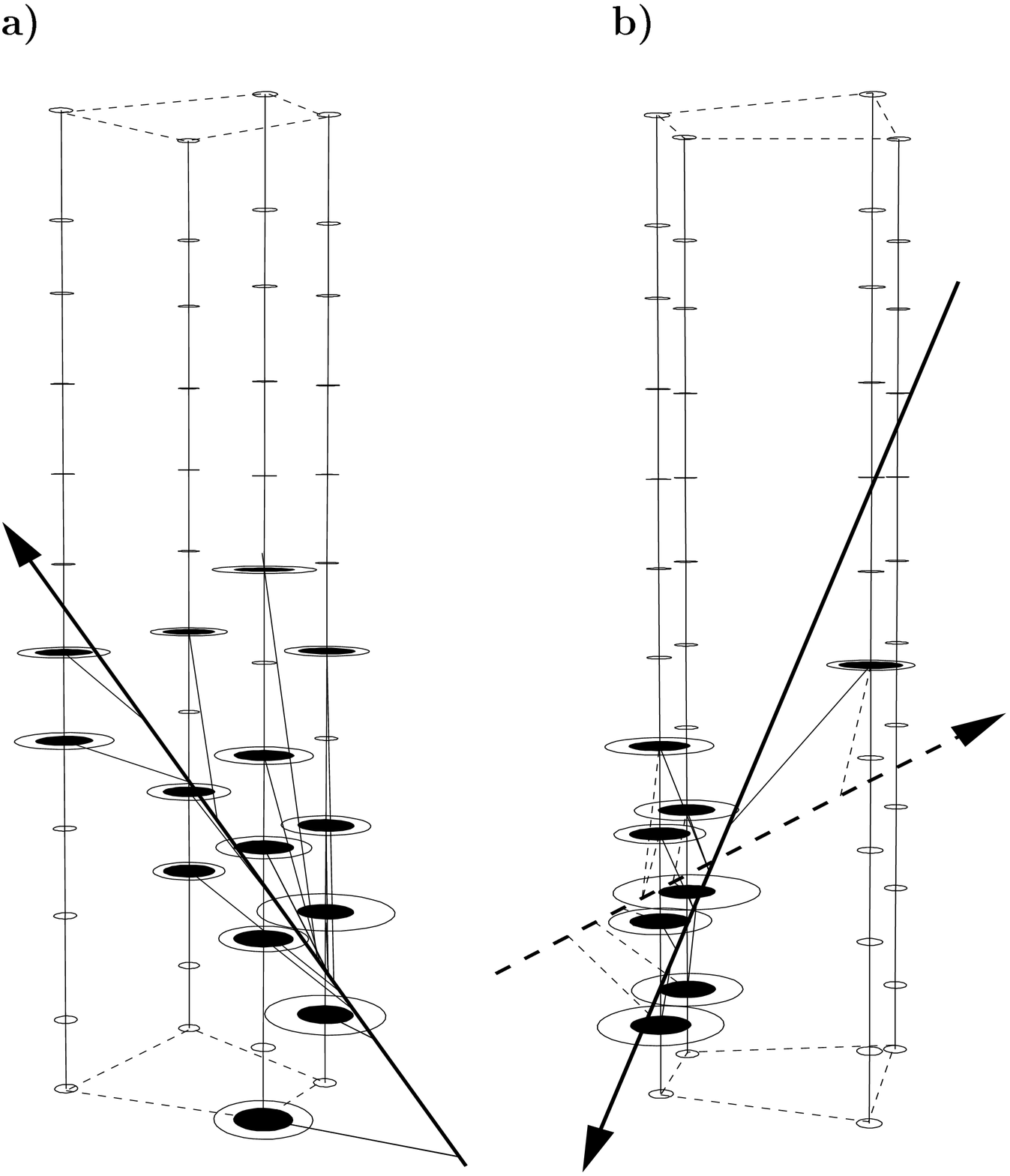,height=16cm}}
\end{figure}

\vspace{0cm}

{\bf Figure 8:} a) - an unambiguous 14-hit neutrino candidate; b) - an 
ambiguous event reconstructed as a neutrino event (dashed line) but with a 
second solution above the horizon (solid line). This event was assigned to the
sample of downward going muons.

\newpage

\strut

\vspace{1cm}

\begin{figure}[h] 
\centering
\mbox{\epsfig{file=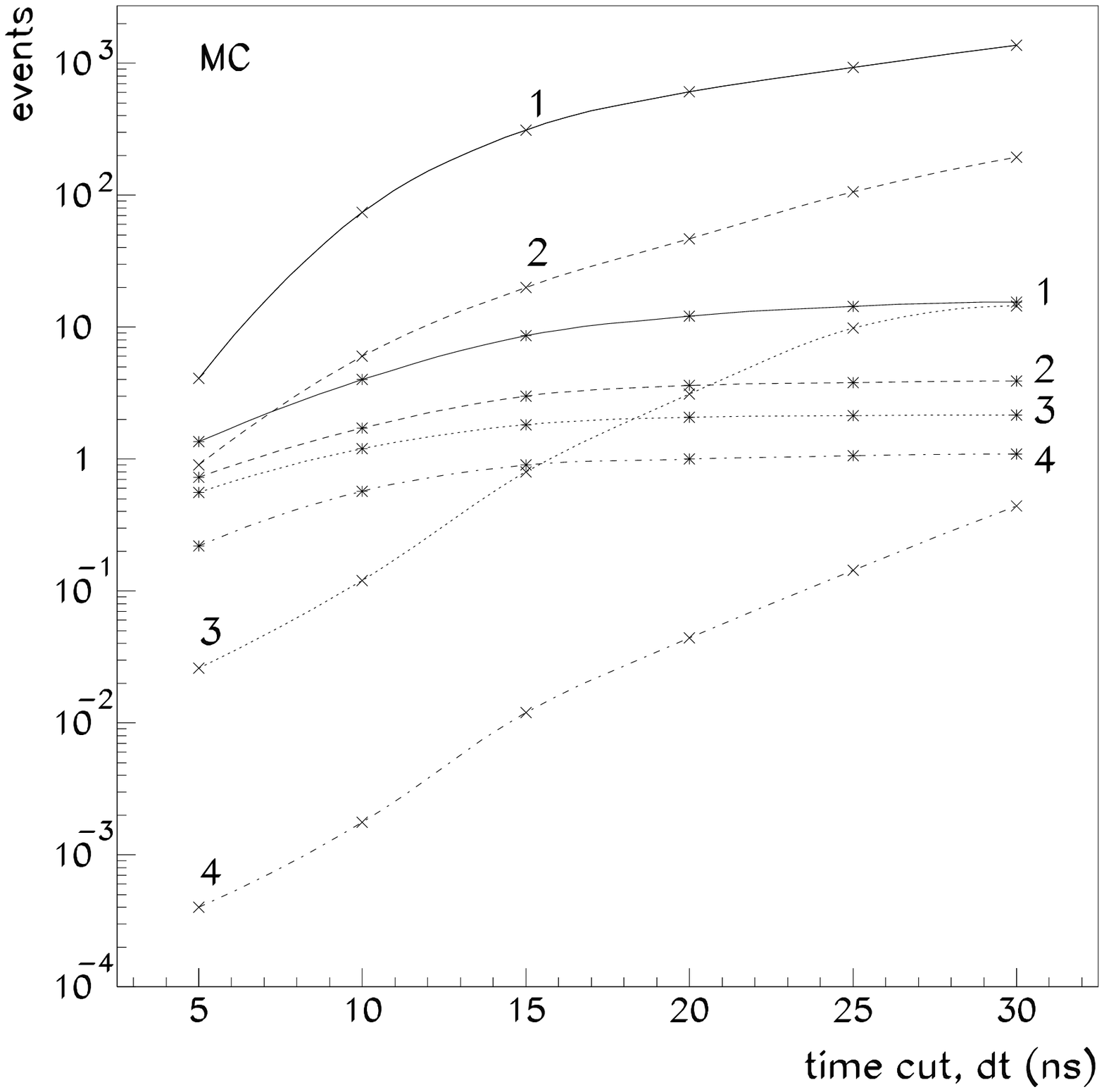,height=16cm}}
\end{figure}

\vspace{0cm}

{\bf Figure 9:} Expected numbers of muons from atmospheric neutrinos 
(asterisks) and background events (crosses) per year vs. time cut $dt$. 
Curves marked 1; 2; 3 and 4 correspond to trigger conditions {\it 1, 1-2, 1-3}
and \mbox{{\it 1-4.}}

\newpage

\strut

\vspace{1cm}

\begin{figure}[h] 
\centering
\mbox{\epsfig{file=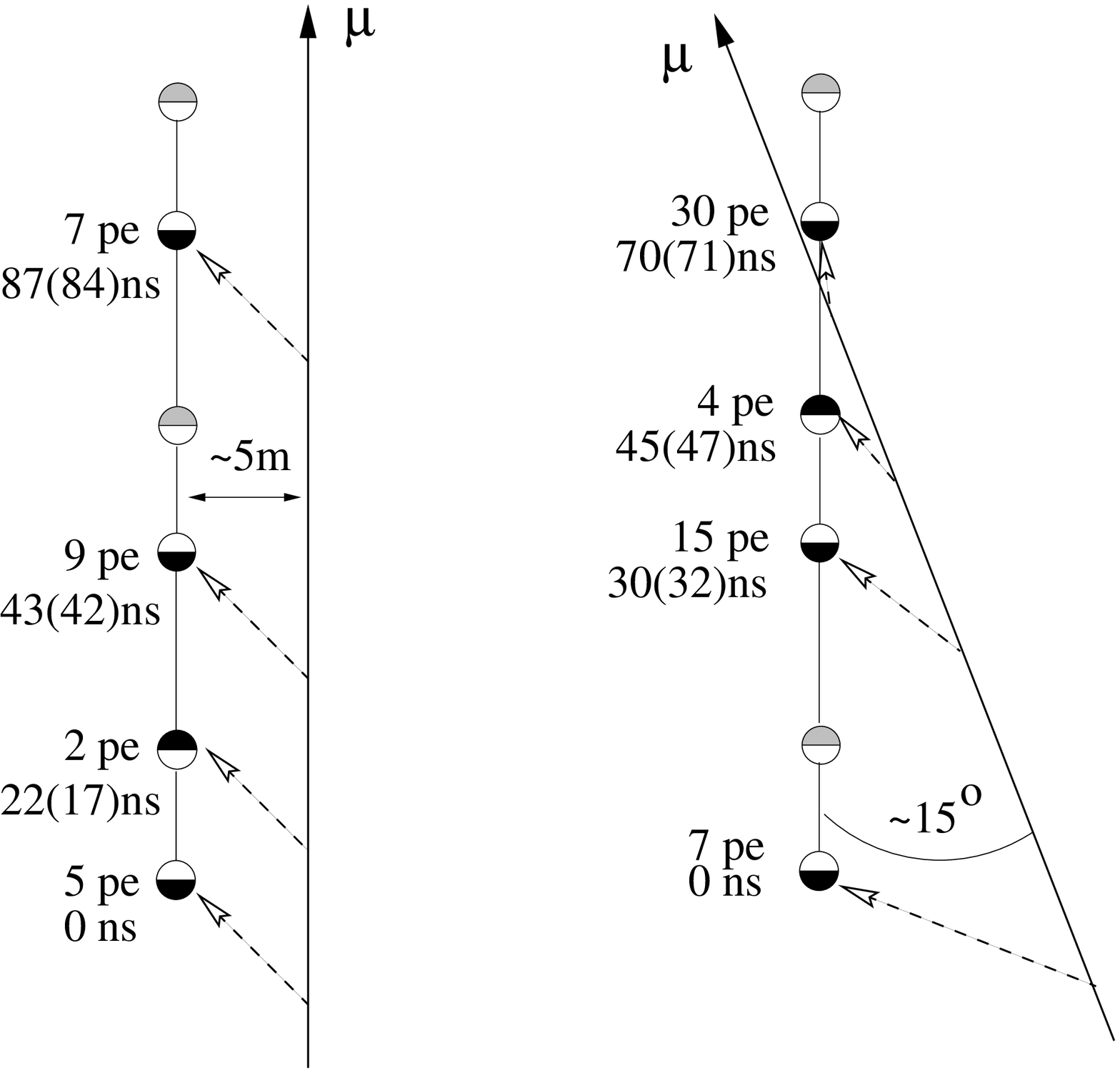,height=14cm,angle=-90}}
\end{figure}

\vspace{1cm}

{\bf Figure 10:} The two neutrino candidates. The hit PMT pairs (channels) are
marked in black. Numbers give the measured amplitudes (in photoelectrons) and 
times with respect to the first hit channel. Times in brackets are those 
expected for a vertical going upward muon (left) and an upward muon passing 
the string under \mbox{$15^o$} (right).

\newpage

\strut

\vspace{0cm}

\begin{figure}[h] 
\centering
\mbox{\epsfig{file=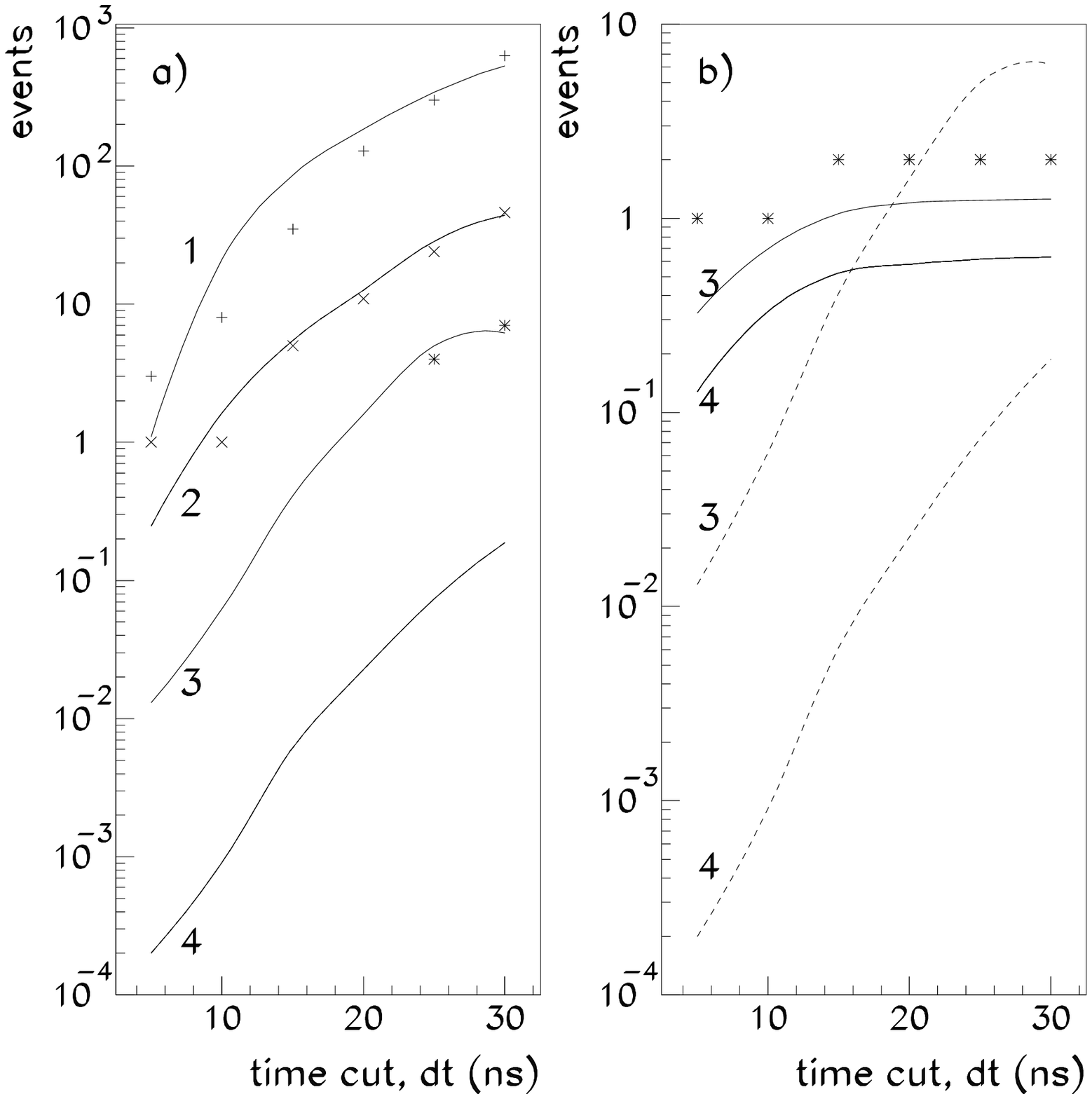,height=16cm}}
\end{figure}

\vspace{1cm}

{\bf Figure 11:} Distribution of experimental sample vs. time cut $dt$. 
Numbers 1; 2; 3 and 4 correspond to trigger conditions {\it 1, 1-2,1-3} and 
{\it 1-4}, respectively. a) - background events: lines present MC expectations 
for different trigger conditions; b) - neutrino candidates: solid and dashed 
lines present MC expectations for upward going  muons generated by atmospheric
neutrinos (not taking into account light scattering in water) and for 
background events.  

\newpage

\strut

\vspace{0cm}

\begin{figure}[h] 
\centering
\mbox{\epsfig{file=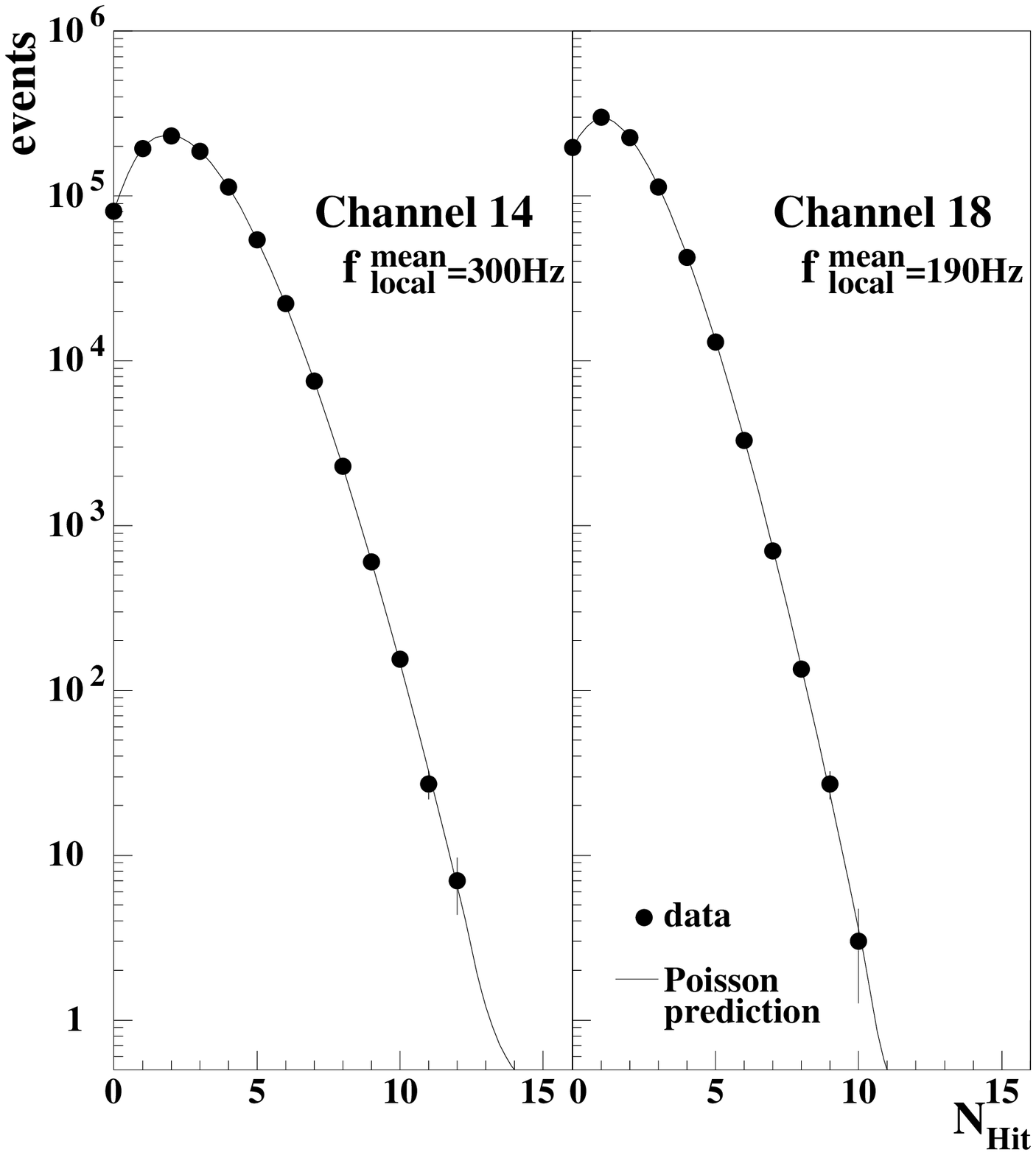,height=16cm}}
\end{figure}

\vspace{1cm}

{\bf Figure 12:} Hit distributions for a time window of 8 msec, as recorded in
a 2 hours test run for channels 14 and 18 of NT-36. Experimental data are 
indicated by points, the curve gives the Poisson prediction.

\newpage

\strut

\vspace{0cm}

\begin{figure}[h] 
\centering
\mbox{\epsfig{file=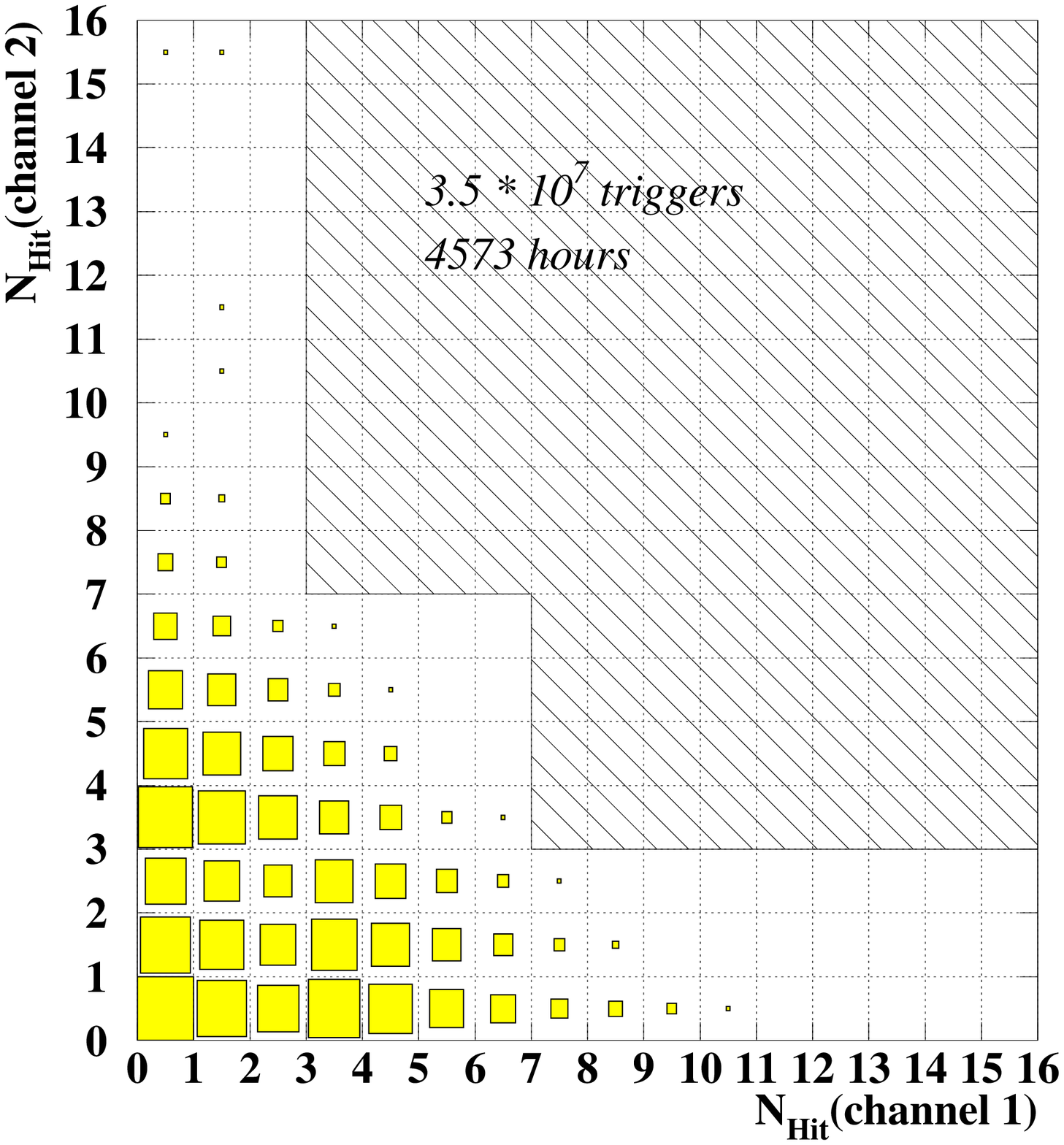,height=16cm}}
\end{figure}

\vspace{1cm}

{\bf Figure 13:} Hit numbers in channel 2 versus channel 1 for  the channels of
all operating $svjaskas$. The shaded region corresponds to the off-line 
trigger condition (see text).

\newpage

\strut

\vspace{0cm}

\begin{figure}[h] 
\centering
\mbox{\epsfig{file=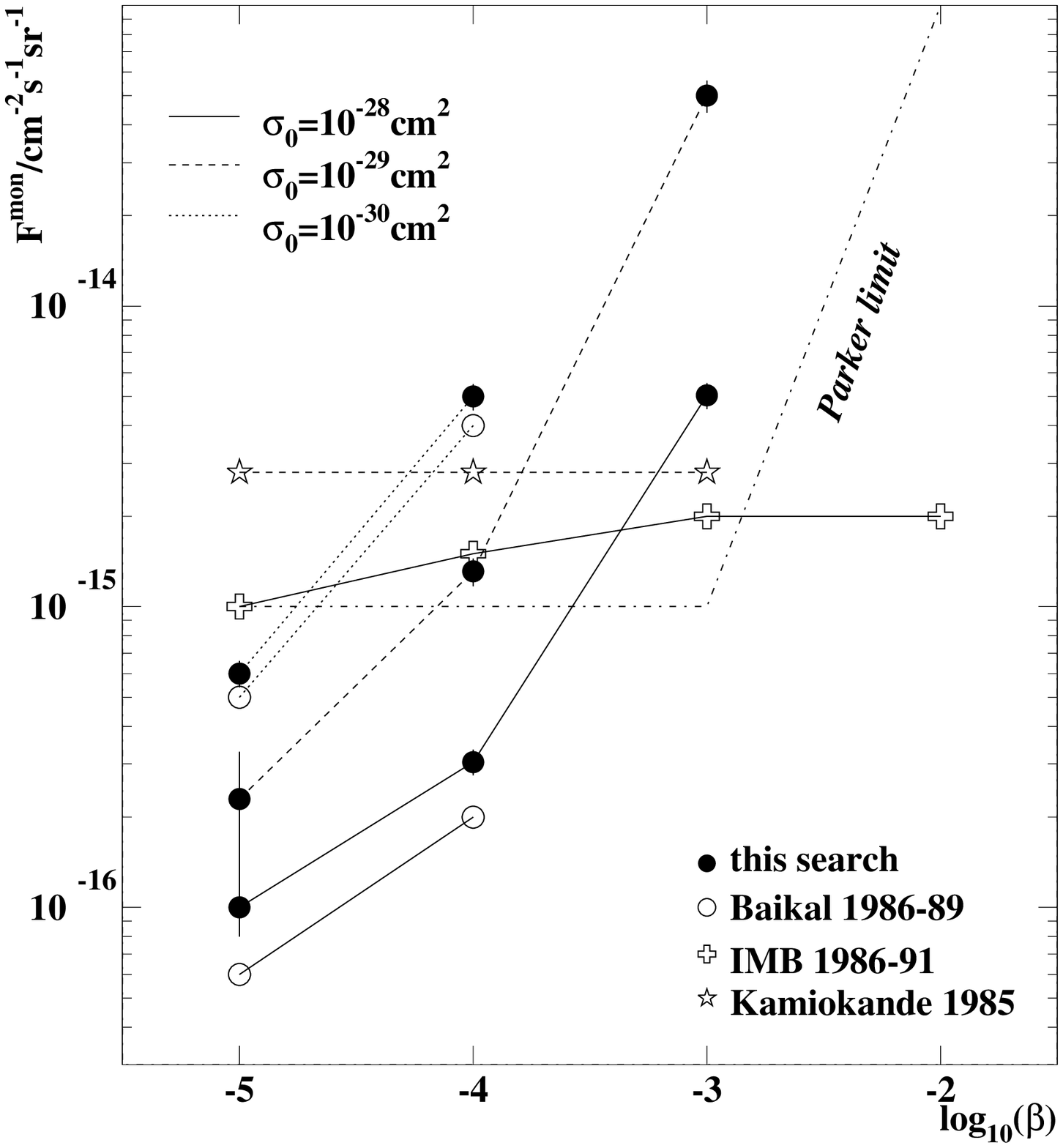,height=16cm}}
\end{figure}

\vspace{1cm}

{\bf Figure 14:} \mbox{Upper limits (90 $\%$ CL)} on the natural flux of 
magnetic monopoles versus their velocity $\beta$, for different parameters
$\sigma_o$ (see text).

\end{document}